\documentclass[12pt,a4paper]{article}
\usepackage{epsfig}
\def\slash#1{\setbox0=\hbox{$#1$}#1\hskip-\wd0\dimen0=5pt\advance
\dimen0 by-\ht0\advance\dimen0 by\dp0\lower0.5\dimen0\hbox
to\wd0{\hss\sl/\/\hss}}
\def\Black{}

\def\Brown{}
\newcommand{\be}{\begin{equation}}
\newcommand{\ee}{\end{equation}}
\newcommand{\bea}{\begin{eqnarray}}
\newcommand{\eea}{\end{eqnarray}}
\newcommand{\dd}{\displaystyle}
\newcommand{\nn}{\nonumber}
\begin{document}
\begin{titlepage}
\null
\begin{center}
\Large\bf \Brown Cottingham formula and the pion electromagnetic
mass difference at finite temperature \Black
\end{center}
\vspace{0.5cm}

\begin{center}
\begin{large}
\vspace*{1cm} {\large M. Ladisa$^{a,b}$, G. Nardulli$^{b}$, S.
Stramaglia $^{b}$\\} \vspace*{1.cm}

\end{large}

$^{a}$ Centre de Physique Th\'eorique, \\ Centre National de la
Recherche Scientifique, UMR 7644, \\ \'Ecole Polytechnique, 91128
Palaiseau Cedex, France\\  \vspace*{0.3cm} $^{b}$ Dipartimento di
Fisica, Universit\`a di Bari, Italy, and \\ Istituto Nazionale di
Fisica Nucleare, Sezione di Bari, Italy
\end{center}

\vspace{0.3cm}

\begin{center}
\begin{large}
\Brown {\bf Abstract}\\[0.5cm]\Black
\end{large}
\parbox{14cm}{We generalize the Cottingham formula at finite ($T\neq 0$) temperature
by using the imaginary time formalism. The Cottingham formula
gives the theoretical framework to compute the electromagnetic
mass differences of the hadrons using a dispersion relation
approach. It can be also used in other contexts, such as non
leptonic weak decays, and its generalization to finite temperature
might be useful in evaluating thermal effects in these processes.
As an application we compute the $\pi^+-\pi^0$ mass difference at
$T\neq 0$; at small $T$ we reproduce the behaviour found by other
authors: $\delta m^2(T)= \delta m^2(0)+\mathcal{O}$$(\alpha T^2)
$, while for moderate $T$, near the deconfinement temperature, we
observe deviations from this behaviour.}
\end{center}

\vspace{1.0cm} \noindent \Black \vfil \noindent \Brown
BARI-TH/99-348\\ July 5th, 1999 \Black
\end{titlepage}

\section{Introduction}
The aim of this paper is to present the generalization of the
Cottingham  formula \cite{Cottingham} to finite temperature (for a
review of field theory at finite temperature and density see
\cite{bellac}). The Cottingham formula allows the evaluation of
the time-ordered product of two hadronic currents between hadronic
states by a Dispersion Relation (DR) and its  main application is
in the evaluation of electromagnetic mass differences of hadrons.
The  use of the Cottingham  formula  in this context predates
Quantum Chromo Dynamics (QCD) \cite{harari}; by the advent of QCD
some modifications of the original formalism were adopted: for
example,
 the problem of the convergence was settled by the use
 of an ultraviolet cut-off $\mu^{2}$ related to the renormalization procedure
 \cite{collins}. Furthermore, the use of effective field theories: chiral perturbation
 theory \cite{donoghue} and   heavy quark effective theory
 \cite{ladisa},  has introduced  well defined theoretical schemes to evaluate the relevant
 Feynman diagrams.

Besides the use in the context of the electromagnetic mass
differences of hadrons, the Cottingham formula has ben also
applied in other fields, mainly related to non leptonic weak
processes: kaon decays \cite{rotondi}, \cite{pham},
$K^0-\bar{K^0}$ mixing \cite{cea}, parity violations in nucleon
interactions \cite{preparata}, $B$-meson processes \cite{paver};
therefore its extension to finite temperature may be of some
interest to analyze the role of the thermal effects in these
physical situations. Among the possible applications we shall pick
up in this paper the $\pi^+-\pi^0$ mass difference at finite
temperature, a subject that has received a continuous attention
both in the past \cite{kapusta} and in more recent times
\cite{manuel}.

\section{Cottingham formula at finite temperature} As already stressed in the introduction, the main
application of the Cottingham formula is for the calculation of
the electromagnetic mass splitting of mesons. The mass shift of
the meson $M$ due to the electromagnetic interaction can be
obtained by computing:
\begin{equation}
\delta m^{2}=\frac{ie^{2}}{2}\int \frac{d^{4}q}{(2\pi
)^{4}}\frac{g_{\mu \nu }}{q^{2}+i\epsilon }T^{\mu \nu }(q,p)\ ,
\label{shift}
\end{equation}
where the hadronic tensor
\begin{equation}
T^{\mu \nu }(q,p)=i\int d^{4}x\ e^{-iqx}\langle M( p)|T(J^{\mu
}(x)J^{\nu }(0)|M(p) \rangle   \label{T-product}
\end{equation}
describes the Compton scattering of a virtual photon of four
momentum $q^\mu$ off the meson $M$ of momentum $p^\mu$ and
$J^{\mu }$ is the electromagnetic current.  The Compton amplitude
can be decomposed in terms of gauge invariant tensors as follows:
\begin{equation}
T^{\mu \nu }(q,p)=D_{1}^{\mu \nu }\ T_{1}(q^{2},\nu )\ +\
D_{2}^{\mu \nu }\ T_{2}(q^{2},\nu )\ ,  \label{decomposition}
\end{equation}
where
\begin{eqnarray}
D_{1}^{\mu \nu } &=&-g_{\mu \nu }+\frac{q^{\mu }q^{\nu }}{q^{2}}\
, \label{decomposition2} \\ D_{2}^{\mu \nu }
&=&\frac{1}{m^{2}}\left( p^{\mu }-\frac{\nu }{q^{2}}q^{\mu
}\right) \left( p^{\nu }-\frac{\nu }{q^{2}}q^{\nu }\right) ,
\nonumber \\ \nu &=&pq\ .  \nonumber
\end{eqnarray}

The Lorentz invariant structure functions $  T_{1}(q^{2},\nu )$
and $  T_{2}(q^{2},\nu )$ depend, in the meson rest frame, by
$|\vec q | $ and $\dd{q^{0}=\frac{\nu }{m}}$, where  ${m}$ is the
meson mass. For fixed $|\vec q | $, the singularities  in the
complex $q^{0}$ plane are placed just below the positive real axis
and just above the negative real axis; therefore one can perform a
Wick rotation to the imaginary axis $q^0=ik^{0}$, without
encountering any singularity. After this transformation, the
integration involves only space-like momenta for the photon:
\begin{eqnarray}
q^{0} &\rightarrow &ik^{0},  \label{new-variables} \\ q^{2}
&\rightarrow &-Q^{2}=-(k_{0}^{2}+|\vec{q}|^{2})\ . \nonumber
\end{eqnarray}

After the change of variables (\ref{new-variables}) one obtains
the Cottingham formula \cite{Cottingham}:
\begin{eqnarray}
\delta m^{2}&=&\frac{e^{2}}{16\pi ^{3}}\int_{0}^{+\infty }\frac{dQ^{2}}{Q^{2}}%
\int_{-\sqrt{Q^2}}^{+\sqrt{Q^2}}dk^{0}\sqrt{Q^{2}-k_{0}^{2}}\times
\nonumber\\ &\times &\left[ -3T_{1}(-Q^{2},i m k^{0})+\left(
1-\frac{k_{0}^{2}}{Q^{2}}\right) T_{2}(-Q^{2},i m k^{0})\right] \
. \label{shift2}
\end{eqnarray}

As discussed in \cite{collins} and  \cite{ladisa},  the  $Q^2$
integration range is cut-off at $Q^2_{max}=\mu^2$.  $\mu$   is the
Quantum Chromo Dynamics (QCD)  renormalization mass scale, at
which both the strong coupling constant $\alpha_s$ and the quark
masses $m_q$ have to be specified and roughly represents the onset
of the scaling behaviour of QCD. As is well known the
renormalization procedure introduces counterterms which  cancel
the infinite contribution induced by virtual particles with
momenta larger than $\mu$: therefore its net effect is analogous
to a cut-off of the $Q^2$ integral at $\mu^2$.
 For approximate calculations  there is a residue smooth dependence on $\mu$,
which in a complete calculation   is exactly canceled by the
$\mu-$dependence of the renormalized quark masses and strong
coupling constant.  Typical values of $\mu$ are in the range of
1-3 GeV, corresponding to the onset of the scaling behaviour of
QCD and to a mass scale significantly larger than  all the
hadronic masses.

We wish now to generalize this formula at finite temperature; we
choose to work in the imaginary time formalism, which corresponds
to substitute the integral over the energies with a discrete sum
over the so called \emph{Matsubara frequencies $~\omega_n~=~2\,
\pi\,  n\,  T$}:
\begin{equation}\label{matsu}
  \int_{-\infty}^{+\infty}d q^0 f(q^0  )
  ~\rightarrow~{~}2\, \pi\, T\sum_{n=-\infty}^{+\infty}f(q^0
  )\Big|_{q^0=~i~\omega_n}~{~}.
\end{equation}
This substitution corresponds to insert in (\ref{shift2})  the
factor
\begin{equation}\label{matsu2}
2\, \pi\, T\sum_{n=-\infty}^{+\infty}\delta(k^0~-~\omega_n)~.
\end{equation}
We therefore obtain, from (\ref{shift2}) and (\ref{matsu2}):
\begin{eqnarray}
\delta m^{2}&=&\frac{\alpha}{4\pi^{2}}\int_{0}^{+\mu^2
}\frac{dQ^{2}}{Q^{2}} \, 2\, \pi\,
T\sum_{n=-[a]}^{+[a]}\sqrt{Q^{2}-\omega_n^2 }\times \nonumber\\
&\times &\left[ -3T_{1}(-Q^{2},\, i\, m \, \omega_n)+\left(
1-\frac{\omega_n^{2}}{Q^{2}}\right) T_{2}(-Q^{2},\, i\, m\,
\omega_n)\right] \ , \label{sshift3}
\end{eqnarray}
where
\begin{equation}\label{a}
a=\frac{\sqrt{Q^2}}{2\pi T}~.
\end{equation}
$[a]$ represents the maximum integer contained in the real number
$a$.

Eq.(\ref{sshift3}) generalizes the Cottingham formula at $T\neq 0$
and is the starting point of the application to the
$\pi^+~-~\pi^0$ mass difference at finite temperature to be
discussed in the following section.

\section{The pion electromagnetic
mass difference at finite temperature\label{pion}} The invariant
amplitudes $T_{1} $  and $T_{2} $ satisfy dispersion relations
(DR) in the $\nu$ variable. The DR for $T_2$ is unsubtracted,
while  $T_{1}$ requires one subtraction\cite{harari},
\cite{donoghue}:
\begin{eqnarray}
T_{1}(q^{2},\nu ) &=&T_{1}(q^{2},0)~+~\frac{\nu\,^{2}}{\pi
}\int_{0}^{+\infty } \frac{d\nu\,^{\prime\, 2}}{\nu\,^{\prime\,
2}}~\frac{{Im}\ T_{1}(q^{2},\nu\,^{\prime })}{\nu\,^{\prime\,
2}-\nu\,^{2}},  \label{DR}
\\ T_{2}(q^{2},\nu ) &=&\frac{1}{\pi
}\int_{0}^{+\infty }d\nu\,^{\prime\, 2}~\frac{ {Im}\
T_{2}(q^{2},\nu\,^{\prime })}{\nu\,^{\prime\, 2}-\nu^{\,2}}\ .
\end{eqnarray}

After the change of variables (\ref{new-variables}) these
equations become:
\begin{eqnarray}
T_{1}(-Q^{2},i\, m\, k^{0}) &=&T_{1}(-Q^{2},0)~-~m^2
k_{0}^{2}\int_{0}^{+\infty }~\frac{ d\nu\,^{\prime\,
2}}{\nu\,^{\prime\, 2}}~\frac{W_{1}(-Q^{2},\nu^{~\prime
})}{\nu\,^{\prime\, 2}+m^2\, k_{0}^{2}}, \label{DR2} \\
T_{2}(-Q^{2},i \, m\, k^{0}) &=&\int_{0}^{+\infty }d\nu\,^{\prime
\, 2}~ \frac{ W_{2}(-Q^{2},\nu\,^{\prime })}{\nu\,^{\prime\,
2}+m^2\, k_{0}^{2}}, \label{DR2bis} \\ W_{i}(q^{2},\nu) &=
&\frac{1}{\pi }~{Im}\ T_{i}(q^{2},\nu)\ ,\ (i=1,2).
\end{eqnarray}

In order to evaluate the DR in the case of the  pion
electromagnetic mass difference  we consider the contribution of
the Born term (the $\pi$ meson itself) and  the $J^P=1^-$
resonances  $\omega $. Other contributions might be in principle
sizeable, however it is well known, since the work by Harari
\cite{harari}, that the DR are sufficiently well convergent at
$T=0$ and the first two polar terms in (\ref{DR}) represent by far
the largest contribution to the electromagnetic mass difference.
This stems from the fact that the $\pi^+ -\pi^0$ mass difference
takes contribution from the Isospin$=2$ mass term, and the time
ordered product of the two electromagnetic currents in this case
has no singularities for $x \to 0$ (or, equivalently, for
$Q^2\to\infty$), differently from other hadronic mass differences,
such as the proton-neutron mass difference, that present a $1/x^2$
light-cone singularity. For the same reason the limit $\mu \to
\infty$ could also be taken in this case.

To compute the different contributions  to (\ref{DR})  we consider
the following matrix elements ($q=p'-p$): \bea <\pi^+(p')~| J_{\rm
em}^{\mu} ~|~\pi^+(p)> &=&
 F(q^2) ~(p + p')^\mu \label{13}\\
<\, \omega(p',\epsilon)~| J_{\rm em}^{\mu} ~|~\pi^{0}(p)> &=& i~
h(q^2) ~\epsilon^{\mu\lambda\rho \sigma} \epsilon^*_\lambda~
q_\rho ~p_\sigma \label{14}
 \eea
 where   $\epsilon_\lambda$
 is the  $\omega$ polarization vector and  $F,~h$ are electromagnetic form factors.
They can be written as follows:
 \bea
F(q^2) &=&\frac{1}{1- q^2/m_V^2  } \label{16}
\\
h(q^2) &=& ~ \frac{h(0)} { 1- q^2/m_V^2} \label{17} \eea $m_V$ is
a mass parameter that  we identify with the $\rho$ mass. Indeed
both form factors can be obtained by assuming $\rho$ dominance.
This hypothesis allows to extrapolate the behaviour at finite
temperature. Indeed at $T\neq 0$, but small, one has \cite{koch}
\begin{eqnarray}
 F(q^2,T)&=&
 \left[1+\frac{2 T^2}{3 f_\pi^2} g_0\left(\frac{m_\pi^2}{T^2}\right)\right]\times\nn
 \\
&& \times\left[
  \frac{g_{\rho\pi\pi}(T)g_{\rho\gamma}(T)}{m^2_\rho}\frac{1}{1- q^2/m_\rho^2
  }~-~\frac{T^2}{4f_\pi^2}g_0\left(\frac{m_{\pi}^2}{T^2}\right)\right]~.\label{19bis}
\end{eqnarray}
This expression takes into account the effective charge of a pion
in thermal medium containing other pions at the equilibrium.  Here
$f_\pi=93$ MeV,
 $g_0(x)$
 is given by \cite{koch}:
\begin{equation}\label{g0}
g_0(x)=\frac{1}{2\pi^2} \int_{0}^{\infty} d y
\frac{y^2}{\sqrt{x^2+y^2} \exp(\sqrt{x^2+y^2})-1)}~,
\end{equation}
while $g_{\rho\pi\pi}(T)$ and $g_{\rho\gamma}(T)$ are coupling
constants for the $\rho\pi\pi$ and  $\rho-$photon vertex
respectively, computed taking into account the effects of the
thermal bath. They are obtained in the soft pion limit, which is
sufficiently accurate for our purposes \cite{koch}.  These effects
(and the related $f_\pi(T)$ dependence) have been computed by
several authors and with different methods (see, for example
\cite{koch},\cite{song} and references therein). The results are:
 \bea
  g_{ \rho \pi \pi}(T)  &= &g_{ \rho \pi \pi }\left[  1-\frac{5 T^2}{12 f_\pi^2}
  g_0\left(\frac{m_\pi^2}{T^2}
  \right)\right]\\
 g_{\rho\gamma} (T) &=&
g_{\rho\gamma}\left(1-\frac{T^2}{12 f_\pi^2} \right)
 \eea
 The values of these constants at $T=0$ are related by the formula
\begin{equation}\label{dom}
\frac{  g_{ \rho \pi \pi} g_{\rho\gamma}}{m^2_\rho}=1~,
\end{equation}
a relation that implements the idea of $\rho$ dominance of the
electromagnetic  form factor.
 It is worth stressing that, independently of the thermal effects
 induced by electromagnetism, the pion mass gets a thermal self
 energy also by the strong interactions with the dilute pion gas. In
 general we shall not consider these effects since they are
 identical for $\pi^+$ and $\pi^0$ and cancel in the difference.
 In one case the thermal self energy acts however as an infrared
 regulator (see below). We quote therefore this
 effect \cite{manuel}:
\begin{equation}\label{manul}
m_{\pi^{+,0}}^2(T)=m_{\pi^{+,0}}^2\left( 1-\frac{ T^2}{6 f_\pi^2}
\right)
\end{equation}

We also observe that, because of the interaction with the thermal
bath, pions can couple directly to the photon without the
intermediate virtual $\rho$ state: this effect is at the origin of
the last term in (\ref{19bis}). This term, which only exists at
$T\neq 0$, gives a non-vanishing contribution to the form factor
for $Q^2\to+\infty$, which might be an artifact of the effective
theory employed to compute it. However, as the $Q^2$ integration
is cut-off at $\mu^2$ and the results depends smoothly on $\mu$,
there is no need to correct for this effect and we shall assume
(\ref{19bis}) as it stands.

 Let us now turn to the form factor
relative to the $\omega$ intermediate state. We put $h(0)=2.6$
GeV$^{-1}$, which can be derived by $\omega\to\pi\gamma$ decay
rate.  As it will be clear in the sequel, a more precise
determination of $h(q^2)$ is not necessary because of the smaller
role played by the $\omega$ intermediate state in the calculation.
For the same reason we omit to include thermal effects in the
vertex $\omega\pi\gamma$ that, to our knowledge, have not yet been
computed.

 Using the matrix
elements and the coupling constants already introduced, we can
calculate the electromagnetic pion mass splitting. To start with,
we consider the $T=0$ case. The contributions of the different
terms to the DR are as follows; the subtraction term $T_1(q^2,0)$
is given by:
\be
T_1(q^2,0) = - 2 F^2(q^2) +
 \frac{m^2 q^2 h^2(q^2)}{\nu_R} ~.
\ee

The two structure functions  $W_{1,\, 2}(q^2,\nu)$ that appear in
(\ref{DR2}) and (\ref{DR2bis})  are given  by:
 \bea W_1(q^2,\nu) &=&+\nu_R\left(m^2q^2-\nu_R^2\right) h^2(q^2)
~\delta\left( \nu^2 - \nu_R^2  \right)
\\
W_2(q^2,\nu) &=& - 2 m^2 q^2 F^2(q^2)~\delta \left(
\nu^2-\frac{q^4}{4} \right) \nn\\ & +& q^2 m^2\nu_R~ h^2(q^2)
~\delta \left(\nu^2 - \nu_R^2\right) \eea where
$\dd{\nu_R=\frac{m_\rho^2-m^2-q^2}{2}}$.
 It is worth stressing that the
Born term  contributes only to $W_{2}(q^2,\nu)$, its contribution
to $T_1$ only being through $T_1(q^2,0)$.

 These expressions can be used to compute the
electromagnetic pion mass difference at $T=0$. The results are
reported in Table I, for $\mu=2$ and $\mu=3$   GeV and for
$\mu=\infty$.  It may be useful to stress that the numerical
results are remarkably insensitive to variations of the cut-off.
They show  show that the pion contribution represents by far the
dominant part of  the mass difference. The small difference
(around $10\%$) between the data and the theoretical result should
be attributed to a few other poles in the dispersion relation,
most notably the $a_1$ resonance. \vskip 0.5 truecm
 \begin{center}
\begin{tabular}{|c|c|c|c|}
  \hline
  {\rm Contribution } & $\mu=~2${\rm~ GeV} & $\mu=~3${\rm~ GeV}& $\mu=~\infty$\\
  $\pi$ &3.56 &3.82 &4.05\\
  $\omega$ & 0.09& 0.11 &0.13\\
  {\rm Total} &3.65  & 3.93 & 4.18\\ \hline
\end{tabular}\vskip 0.2 true cm
 {{\bf Table I.} Contributions to $\delta m $ in MeV;  $\delta m_{exp} =4.6$ MeV.}
\end{center}
\vskip 0.4 true cm
 In the chiral limit $m_\pi=0$ we get, for
$\mu=\infty$:
\begin{equation}\label{chirres}
  \delta m^2=\frac{3\, \alpha\, m_\rho^2}{4\, \pi}~,
\end{equation}
which gives,  numerically, $\delta m=m_{\pi^+} - m_{\pi^0}=3.8$
MeV, to be compared  to the the experimental result $\delta
m_{exp}=4.6$ MeV. We can also compare it to the the result of Das
et al. \cite{das}:
\begin{equation}\label{chidas}
  \delta m^2=\frac{3\, \alpha
  \, m_\rho^2}{4\, \pi}\frac{f_\rho^2}{f_\pi^2}\ln \Big( \frac{f_\rho^2}{f_\rho^2-f_{\pi}^2} \Big)
\end{equation}
that is obtained using the chiral limit and the Weinberg sum
rules. The extra factor in (\ref{chidas}) as compared to
(\ref{chirres}) is given by $~\dd{\frac{f_\rho^2}{f_\pi^2}\ln\Big(
\frac{f_\rho^2}{f_\rho^2-f_\pi^2}\Big) }~$ and is numerically
equal to $1.24$ for $\dd{\frac{f_\rho}{f_\pi}} =1.66$.

 Let us now consider the analogous calculation  at $T\neq 0$. We
 shall use in (\ref{sshift3})
 \bea
 T_1(-Q^2,i\, m\, \omega_n)&=& -2 F^2(-Q^2,T) \nn \\
& -& m^2 h^2(-Q^2)\Big[ \frac{2 Q^2}{Q^2+m^2_\omega-m^2}-
 \frac{\omega_n^2(\nu_R^2+m^2Q^2)}{\nu_R(\nu_R^2+m^2\omega_n^2
 )}\Big] \\
 T_2(-Q^2,i\, m\, \omega_n)&=& 8 m^2 F^2(-Q^2,T)\frac{ Q^2-m^2+m^2(T)}{
 \Big((Q^2-m^2+m^2(T)\Big)^2+4m^2\omega_n^2}+
 \nn \\
&-& m^2h^2(-Q^2)\frac{ Q^2 \nu_R }{\nu_R^2+m^2\omega_n^2
 }~,
 \eea
where $m^2(T) $ is given in (\ref{manul}). By inserting these
results in eq.(\ref{sshift3}) we can compute numerically the sums
over the discrete energy. The numerical results will be discussed
in the next Section.

\section{Electromagnetic mass difference in the chiral limit}
In the chiral limit the expression for the electromagnetic pion
mass difference looks remarkably simple:
\begin{equation}\label{dmchi}
  \delta m^2(T) =\frac{6\alpha}{4\pi^2}\int_{0}^{\mu^2}\frac{d
  Q^2}{Q^2}F^2(-Q^2,T)~2\pi~T\sum_{n=-[a]}^{+[a]}\sqrt{Q^{2}-\omega_n^2
  }~.
\end{equation}

 The behaviour of $ \delta m^2(T) $ at small T is as follows:
\begin{equation}\label{om1}
 \delta m^2(T) = \delta m^2 +\lambda T^2~,
\end{equation}
where $\delta m^2=  \delta m^2(T=0)$ and $\lambda $  is a
coefficient.

The absence of the term linear in $T$ can be proved explicitly by
performing an asymptotic expansion in $1/T$ of $  \delta m^2(T)$.
The term independent of $T$ is given by
\begin{eqnarray}
     \lim_{T\to 0} \delta m^2(T) &=&
  \frac{6 \alpha}{4\pi^2}\int_{0}^{\mu^2}d Q^2 F^2(-Q^2)\lim_{a\to \infty}\frac{1}{a}
  \sum_{n=-[a]}^{+[a]}\sqrt{1-\frac{n^2}{a^2}  }=\nn \\
  &=& \frac{6\alpha}{4\pi^2}\int_{0}^{\mu^2}d Q^2 F^2(-Q^2)\int_{-1}^{+1} d x \sqrt{1-
  x^2}=\nn \\ &=&\frac{3\alpha}{4\pi}\int_{0}^{\mu^2}d Q^2 F^2(-Q^2)
  ~,
\end{eqnarray}
which coincides with  $\delta m^2$, i.e. with  the result obtained
by putting directly $T=0$ from the very beginning.   As to the
coefficient of the term in $T^2$, let us write it as follows:
\begin{equation}\label{l1l2}
  \lambda=\mathcal{L}_1~+~\mathcal{L}_2~,
\end{equation}
as it arises from two different sources. The first term,
$\mathcal{L}_1$,  is obtained by expanding $F^2(-Q^2,T)$ in $T$
and taking into account that $g_0(0)=1/12$:
\begin{equation}\label{ft}
F^2(-Q^2,T)= F^2(-Q^2)-\frac{T^2}{8
f_\pi^2}\Big[F^2(-Q^2)+\frac{F(-Q^2)}{3} \Big]+\mathcal{O}(T^2)~;
\end{equation}
from this equation one gets the following contribution to $ \delta
m^2(T)$
\begin{equation}\label{primo}
 \delta m^2(0)~\Big[1-\frac{T^2}{8 f_\pi^2}\Big(1+ \frac{m^2_\rho+\mu^2}{3\mu^2}
 \ln\frac{m^2_\rho+\mu^2}{m^2_\rho}  \Big)  \Big]~,
\end{equation}
and, therefore,
\begin{equation}\label{l1}
\mathcal{L}_1=-\frac{T^2}{8 f_\pi^2} \delta m^2(0)~\Big(1+
\frac{m^2_\rho+\mu^2}{3\mu^2}
 \ln\frac{m^2_\rho+\mu^2}{m^2_\rho}  \Big) ~.
\end{equation}
 The second term, i.e. $\mathcal{L}_2$,  is obtained by putting  \begin{equation}\label{ft2}
F^2(-Q^2,T)= F^2(-Q^2)
\end{equation}
and computing
\begin{equation}\label{om8}
  \mathcal{L}_2= \lim_{T\to 0}~ \frac{6\alpha}{4
  \pi^2T^2}\int_{0}^{\mu^2}d Q^2 F^2(-Q^2)\Big[\frac{1}{a}\sum_{n=-[a]}^{+[a]}\sqrt{1-\frac{n^2}{a^2}}
            -\frac{\pi}{2}  \Big]~,
\end{equation}where, as before, $\dd{a=\frac{\sqrt{Q^2}}{2\, \pi\,
T}}$. We have been unable  to evaluate analytically this limit;
however, when calculated  numerically, for $\mu\geq 0.5-1$ GeV,
the limit is basically independent of $\mu$ and within a few
percent is given by:
\begin{equation}\label{l2}
 \mathcal{L}_2\approx \pi\alpha~.
\end{equation}
Putting the two contributions together we get
\begin{equation}\label{secondo}
  \delta m^2(T)\approx \delta m^2(0)~\Big[1-\frac{T^2}{6 f_\pi^2}\Big(\frac{3}{4}+
  \frac{m^2_\rho+\mu^2}{4\mu^2}
 \ln\frac{m^2_\rho+\mu^2}{m^2_\rho}  \Big)  \Big]~+~\pi\alpha T^2~.
\end{equation}
 We observe however that
the behaviour of the last term in eq. (\ref{secondo}) holds only
at small $T$ ($T\le 100$ MeV). For $\dd{T\geq \frac{\mu}{2\pi}}$,
its $T$-dependence would be linear:
\begin{equation}\label{largeT}
\frac{3\alpha T}{\pi}\int_{0}^{\mu^2}\frac{d
Q^2}{\sqrt{Q^2}}F^2(-Q^2)~,
\end{equation}
but, given the range of the possible  values of $\mu$, such a
behaviour cannot be reached, as the deconfinement process should
take place at much  smaller value of $T$. For $T>100$ MeV
numerical deviations from the behaviour (\ref{secondo}) are
therefore expected, as we will discuss below.

We can compare these results with those obtained by the authors of
 \cite{manuel} in the framework of chiral perturbation theory at
 finite temperature within the hard thermal loop approximation:

\begin{equation}\label{terzo}
  \delta m^2(T)=\delta m^2(0)~\Big[1-\frac{T^2}{6 f_\pi^2} \Big]~+~\pi\alpha T^2~.
\end{equation}
A part from the difference in $\delta m^2(0)$ discussed above, we
get a small deviation also in the correction $\propto T^2$:
whereas the term $+\pi \alpha T^2$ is identical, the remaining
part $\propto T^2$ differs by the factor
\begin{equation}\label{diff}
 \frac{3}{4}+
\frac{m^2_\rho+\mu^2}{4\mu^2}
 \ln\frac{m^2_\rho+\mu^2}{m^2_\rho}~,
\end{equation}
whose numerical value is $1.3 $ at $\mu=2$ GeV  and $1.5$ at
$\mu=3$ GeV. This difference arises from the different treatment
of the virtual photon effect: in particular the logarithmic $\mu$
dependence in (\ref{secondo}) arises from the direct photon
coupling to pions, which is allowed by the pion electromagnetic
form factor at $T\neq 0$, as given by (\ref{19bis}). In particular
if one  assumes that also this direct coupling, i.e. the last term
proportional to $T^2$ in (\ref{19bis}), is multiplied by the
factor $\dd{\left(1-\frac{q^2}{m^2_\rho}\right)^{-1}}$, then the
same result as in \cite{manuel} is obtained.

Let us now turn to  the numerical results of our analysis. In Fig.
1 we report two curves. The solid line is the result of the full
numerical analysis  contained in Section \ref{pion}, while the
dashed line represents $\delta m^2 (T)$ as computed by the
approximate formula (\ref{secondo}). Both curves are obtained at
$\mu=3$ GeV. The comparison  shows that:
\par\noindent
1) For small $T$, the small deviation ($<7\%$) is mainly due to a
chiral symmetry breaking term which is taken into account in the
full calculation (solid line) and not in the approximate
calculation (dashed line). This extra term is part of the pion
pole contribution to the dispersion relation (the $\omega$
contribution is always very tiny).
\par\noindent
2) For large $T$, i.e. $T> 100 $ MeV, the difference  is mainly
due to the fact that the approximation (\ref{secondo}) gets worse
with the increasing temperature; in particular the positive
contribution, i.e. the last term in (\ref{secondo}), is quadratic
only for $T\leq 100 $ MeV, while for $T> 100 $ MeV, it increases
more slowly and the effect of the negative term remains
unbalanced. For $T=200$ MeV this correction is of the order of
$80\%$.

All these results are almost independent of $\mu$ in the range
$2-3$ GeV.

\section{Conclusions} We have extended the
Cottingham formula at finite temperature. This formula allows the
computation of time ordered products of two  currents between
hadronic states; even though it is generally applied to the
calculation of electromagnetic mass differences of hadrons, its
possible range of applications is wider and includes  the
evaluation of matrix elements of other products of currents such
as those occurring in the analysis of weak non leptonic decays.
Therefore its extension to $T\neq 0$ may permit the study of
finite temperature effects also for weak hadronic processes. As an
application, we have considered  the finite temperature
$\pi^+-\pi^0$ electromagnetic mass difference; the use of the
Cottingham formula leads to results similar to those reached by
chiral perturbation theory  for small temperatures $<100$ MeV. We
have also computed deviations due to chiral symmetry breaking,
which are of the order of $7\%$ or less, and corrections to the
hard thermal loop approximation whose role is relevant for $T\geq~
100$ MeV. \vskip 0.5 truecm \noindent{\bf Acknowledgements}\vskip
0.2 truecm \noindent We thank M.Pellicoro for useful discussions.
One of us (M.L.) expresses his gratitude to T. N. Pham for his
encouragements and hospitality.

\newpage
\begin{center}
\begin{large}
{\bf Figure caption}
\end{large}
\end{center}
\par\noindent
Fig. 1. The difference $\delta m^2=m^2_{\pi^+}-m^2_{\pi^0}$ as a
function of the temperature. The solid line is the result of the
full calculation (with a value of the ultraviolet cut-off $\mu=3$
GeV), the dashed line gives the result in the chiral  and  small
temperature limit.

\begin{figure}[t]
\begin{center}
   \includegraphics*[width=14 cm,height=14 cm]{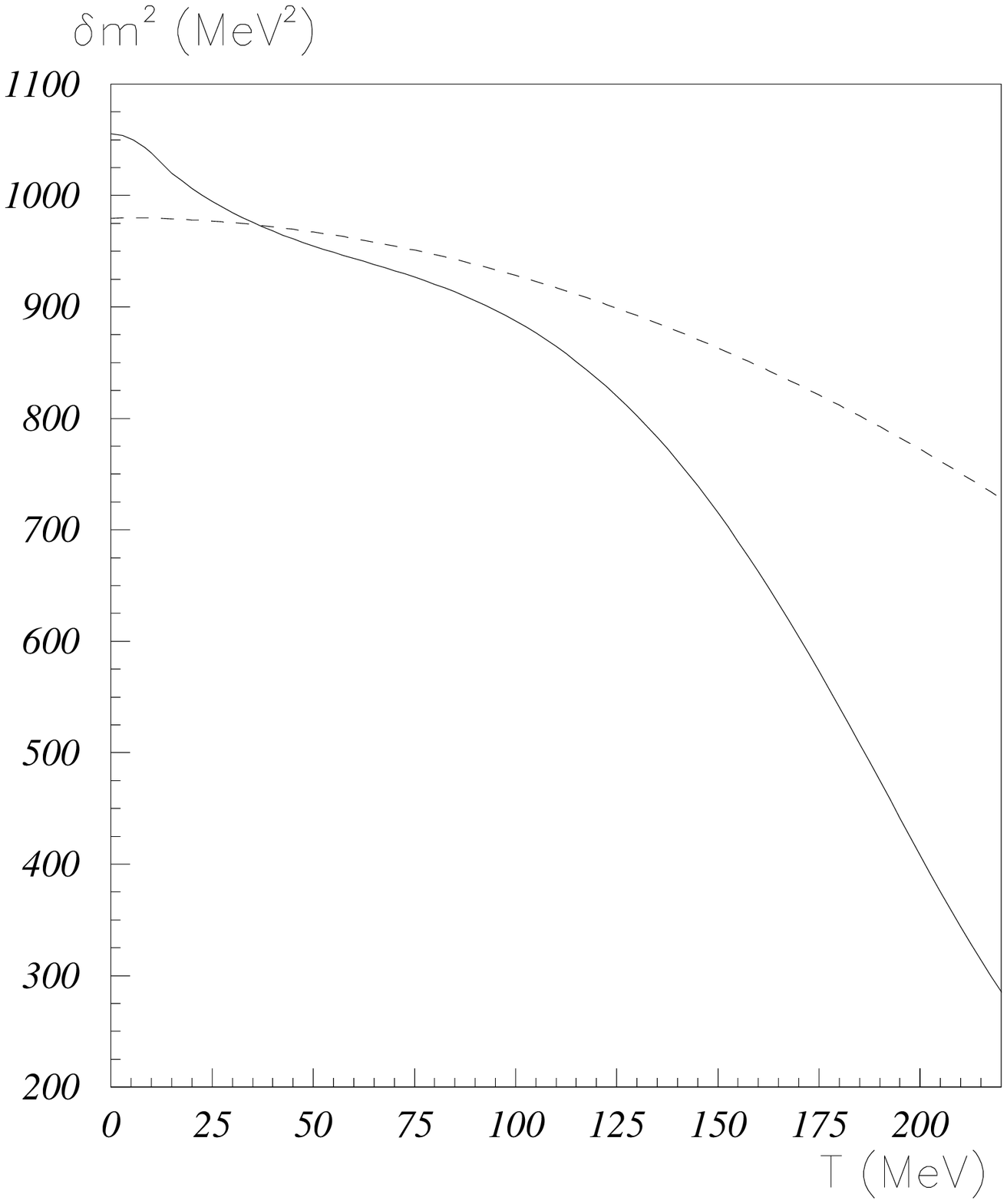}\\*
\label{f:Fig. 1.}
\end{center}
\end{figure}

\end{document}